\documentclass[preprint]{aastex}

\usepackage{lineno}
\shorttitle{%
Simulations of Global Solar Convection with the CHORUS++ Code
}%
\shortauthors{Hayashi et al.}
\begin{document}
\title{%
Simulations of Global Solar Convection with a Fully Compressible CHORUS++ Code
}%
\author{Keiji Hayashi\altaffilmark{1}}\email{keiji.hayashi@njit.edu}
\author{Alexander G. Kosovichev \altaffilmark{1}}%
\author{Chunlei Liang \altaffilmark{2}}
\altaffiltext{1}{New Jersey Institute of Technology, Newark, NJ, U.S.}
\altaffiltext{2}{Clarkson University, Potsdam, NY, U.S.}
\begin{abstract}
Fluid-dynamics simulations of global solar convection are a critically important tool for assessing the dynamics of the solar interior.
However, simulation studies with a fully compressible hydrodynamics code are not yet common.
The CHORUS++ code solves robustly and efficiently the fully compressible hydrodynamics equations using a compact local spectral method and semi-unstructured grid system.
Using the CHORUS++ code, we simulate the solar interior plasma flows from 0.7 to 0.99 of the solar radius using the values of the solar total luminosity and the sidereal rotation rate.
In this paper, we analyze the simulated global flow structures before the statistically stable state
to assess the compressibility of the plasma flows obtained with the fully compressible hydrodynamic code.
The divergence of mass flux with the compressible model is overall small in the examined state before reaching the equilibrium state of the convection, which implies that the differences between the fully compressible flows and those obtained with the anelastic, incompressible, or linear-equation models are small in the simulated inner part of the convection zone.
Although a fully-relaxed stationary convection has not been achieved yet in the examined state, the model qualitatively reproduces the solar-type differential rotation.
Simulations for longer periods are needed to achieve the system relaxation state.
This work, assessing the early phase of the hydrodynamic evolution of solar convection, is our first step toward a better understanding of the nature of solar convection and the dynamo processes.
\end{abstract}
%
%
\section{Introduction}\label{sctintro}
Helioseismic observations from space using the Michelson Doppler Imager (MDI) on board the Solar and Heliospheric Observatory (SOHO) \citep{Scherrer1995} and the Helioseismic and Magnetic Imager (HMI) on board the Solar Dynamics Observatory (SDO) \citep{Scherrer2012}, as well as from the ground-based Global Oscillations Network Group (GONG) \citep{Harvey1996}, revealed the solar internal rotation and its variations with the solar cycle in great detail. It has been established that the latitudinal differential rotation known from observations of the solar surface extends through the whole convection zone, with two distinct regions, characterized by high radial gradients of the rotation rate, the tachocline at the bottom, and the near-surface shear layer at the top. In the bulk of the convection zone, the solar rotation rate has a complicated radial and latitudinal structure with iso-rotational lines forming conical surfaces \citep{Schou1998}. The observations also revealed cyclic variations in the solar rotation, creating a system of zonal flows  (`torsional oscillations'), migrating from mid-latitudes towards the equator and the polar regions \citep{Howe2000}. In addition, local helioseismology methods provided measurements of the meridional circulation and properties of subsurface convection (for a review of recent advances in global and local helioseismology, see \citet{Kosovichev2025} and references therein).    

The observed properties of the solar internal rotation and its variations have not been fully reproduced by theoretical models.
Nevertheless, numerical simulations have provided several key insights into the long-standing question of how plasma flows in the solar convection system interact with rotation and contribute to the 11-year solar-cycle variations of the magnetic field.
For example, the anelastic hydrodynamics models \citep[e.g.,][]{Miesch00,Elliott00} and those coupling with the magnetic field \citep[e.g.,][]{Brun04} successfully retrieved several stellar convection structures, such as the solar-like differential rotation (i.e. the faster rotations at low-latitude regions and slower rotations at the high-latitude regions) under some constraints, the meridional circulation (i.e. poleward flows near the surface and counterflows in the deep convection zone) and the laminar flow patterns when the stellar rotation rate is set high.
The description of the mean-field MHD approach \citep[e.g.,][]{Pipin24} offers a large-scale macroscopic view of azimuthally averaged flows and essential three-dimensional dynamo processes.
The influence of the tachocline and the near-surface magnetic field has been studied by means of numerical implicit Large-Eddy Simulations (ILES) \citep[e.g.,][]{Guerrero16}.
The observation data analysis quantifies the near-surface plasma motions that manifest various forms of global flows, such as meridional flows and convective flows of different temporal and spatial scales \citep[][]{Upton12,Hathaway13}. A notable novel approach is simulations of hydrodynamics (HD) and magnetohydrodynamics (MHD) with the sound-speed reduction method \citep[e.g.,][]{Hotta14,Hotta22}, where the mass continuity equation is modified to reduce the sound speed and hence increase the simulation time step.
This approach has proved efficient, in particular, in parallel computing because the simulation scheme can be highly localized and compact, while
the basic nature of the MHD (or HD),
such as the time-dependent nonlinear behavior of the hyperbolic equation system in multi-dimensional space, and the conservation of mass, momentum, and energy,
is nearly fully taken into account.

As shown in simulations of the interactions between the magnetic field flux and the plasma in the solar convection zone \citep[e.g.][]{Fan03}, compressibility may play a key role in the solar convection system.
Despite limiting the scope to the HD case, density variations play a primary role in various processes, such as driving and maintaining the turbulent flows in the convection system, particularly in the near-surface rotational shear layer.
The material cooled on the solar surface must be heavier because of the local pressure balance, and ultimately, the deviation becomes large enough to break the fragile local hydrostatic balance and start to flow downward.
These nonlinear processes are treated correctly when and only when simulated with fully compressible models
like those for near-surface models
\citep[e.g.,][]{Kitiashvili2023}.

Several factors have prevented the simulations from using the actual solar parameters.
When the actual values of the solar luminosity and rotation rate are used, the evolution from the (usually semi-steady) initial state to the global convective flows will be slow. Hence, it is computationally expensive to simulate the solar convection zone with the actual solar parameters.
A simulation that includes a near-surface region is challenging because of the very steep gradients of plasma density, temperature, and pressure. So far, it has only been done in local rectangular domains.
It requires setting up small grid sizes in the radial direction to treat steep gradients, which requires us to adopt substantially smaller time steps in the ordinary explicit HD or MHD simulation model, leading to increased computational costs.

In part to mitigate these computational difficulties,
a simulation model for solar convection had to apply
a different set of assumptions and simulation parameters to define the simulated situations.
A widely used set of parameters includes larger values of the stellar luminosity 
and/or
the rotation rate than the actual solar parameters
\citep[e.g.,][]{Wang15},
which enhances the growth of convection and hence requires fewer computational resources.
Besides the parameter modification, the assumption of incompressible flow (i.e., $\nabla\cdot(\varrho\vec{V})=0$) or the anelastic model \citep[][]{Gough69} are often used, in part, to achieve computational stability.
The simulations with these assumptions successfully find several important aspects of the solar convection zone.
Nonetheless, to
assess
the difference and/or similarity between the fully compressible flows and the anelastic or incompressible flows,
it is critically important to conduct the simulation of the global solar convection zone with a fully compressible code and with the actual solar parameters.

CHORUS++ \citep[]{Chen23} is an improved version of the CHORUS model \citep[Compressible High-Order Unstructured Spectral differencing][]{Wang15,Wang16}.
CHORUS has two notable features: the unstructured grid system and the spectral differencing method. CHORUS++ employs a cubed-sphere meshing algorithm \citep[]{Ronchi_1996} and improves CHORUS by replacing iso-parametric mappings with transfinite mappings.
As a result, CHORUS++ demonstrated sufficient numerical stability using the 6th- and 7th-order spectral difference methods \citep[][]{Chen23}.
The simulated volume is covered by six subdomains without any overlaps or gaps \citep[e.g.,][]{Feng10}; hence, only flux points on a two-dimensional surface are needed to consider for exchanging physical information between the subdomains.
The spectral differencing method is a subset of the flux reconstruction/correction procedure \citep[e.g.][]{Kopriva_1996,Sun_2007,Huynh07,Huynh09}.
Each subdomain is discretized with a semi-unstructured spherical grid assignment to avoid severe CFL (Courant-Friedrichs-Lewy) condition.
The CHORUS++ code has been demonstrated to achieve improved computational efficiency over the original CHORUS code on distributed-memory CPU systems \citep[]{Chen23} to solve the stellar convection system robustly using a higher-order spatial discretization.

In this article, we use the CHORUS++ code with a few modifications in the simulation setting: the total stellar luminosity is set to the solar luminosity, the rotation rate is set to the solar mean sidereal rotation, and the inner and outer boundary surfaces are set at 0.70 $R_\sun$ and 0.99 $R_\sun$, respectively.
In particular, by setting the outer boundary surface at 0.99 $R_\sun$, we can include the regions with very steep radial gradients of the plasma temperature and density.
With the outer boundary surface close to 1.0 $R_\sun$, the simulations can consider the regions with wider ranges of plasma quantities
(i.e., the contrast of the plasma values at the bottom and top boundaries),
and we can assess the differences and similarities between a fully compressible flow system
and non-fully-compressible systems.

Ideally, the analysis of the simulation results would be conducted for the states of the statistically stable convective system 
after the onset of the convection and an overshoot saturation of the total kinetic energy.
However, the computation, even with the CHORUS++, requires a substantial amount of computation when the top boundary surface is set close to 1 $R_\sun$, because the grid points in the radial direction must be assigned with a smaller grid size to capture and handle the steep gradients of the plasma quantities at $r \lesssim 1 R_\sun$.
Therefore, in this article, we use the simulated data before reaching the statistically stable convection state.
Instead of examining the statistically stable convective system, the analysis is focused on the convection structures in the early phase of convection,
in which the net convective kinetic energy grows exponentially in time, saturates, and starts seeking the stable state.
The analyzed quantities here include the divergence of the mass fluxes to examine the degree of compressibility in the fully compressible system.

This article is organized as follows.
In Section \ref{sctmethod}, the CHORUS++ code and the simulation setting for this study are given.
In Section \ref{sctresults}, the simulation results are shown, and several properties of the derived plasma flows in the solar convection zone, such as the flow patterns and the divergence of the mass flux, are examined.
In Section \ref{sctsummary}, the summary and conclusions are given.

\section{Simulation Model}\label{sctmethod}
CHORUS++ \citep[][]{Chen23} uses the explicit five-stage and third-order Runge-Kutta method of stability preservation (SSPRK (5,3)) \citep[][]{Ruuth05} to solve the set of equations of a fully compressible fluid in the rotating reference frame at a constant rotation rate ($\vec{\Omega_o}$),
\begin{linenomath}\begin{eqnarray}
{\partial\varrho\over\partial t}
&=&
-\nabla(\varrho\vec{U}),
\\
{\partial(\varrho\vec{U})\over\partial t}
&=&
-\nabla\left(\varrho\vec{U}\otimes\vec{U}+P{\cal I}\right)
+\nabla\left(\bar{\bf\tau}\right)
+\varrho\vec{g}-2\varrho\vec{\Omega_o}\times\vec{U},
\\
{\partial{\cal E}\over\partial t}
&=&
-\nabla\cdot\left(\vec{U}({\cal E}+P)\right)
+\nabla\left(\vec{U}\cdot\bar{\bf\tau}-\bar{{\bf q}}\right)
+\varrho\vec{g}\cdot\vec{U},
\end{eqnarray}\end{linenomath}
where
$\varrho$, $\vec{U}$, and ${\cal E}$ are the density, velocity vector, and total energy.
Here, we use the notations of the unit tensor (${\cal I}$) and the tensor product ($\vec{U}\otimes\vec{U}$).
Under the ideal gas assumption, the gas pressure $P$ is given as
$P=\varrho{\cal R}T$ with the gas constant ${\cal R}$, and the total energy is given as
${\cal E}=\displaystyle{\frac{P}{\gamma-1}+{1\over 2}\varrho U^2}$,
with the specific heat ratio $\gamma(=5/3)$.
As in \citet{Chen23}, the centrifugal force is not included in this study.

The shear stress tensor $\bar{\bf\tau}$ is given as
\begin{linenomath}\begin{equation}
\bar{\bf\tau}
=\mu \left(\nabla\vec{U}+(\nabla\vec{U})^T+\lambda(\nabla\cdot\vec{U}){\cal I}\right)
\end{equation}\end{linenomath}
with the dynamic viscosity $\mu=\nu\varrho$, where $\nu$ is the kinematic viscosity coefficient,
and $\lambda=-(2/3)\mu$.
The heat flux ${\bf q}$ is given as
\begin{linenomath}\begin{equation}
{\bf q}=-\kappa\varrho T\nabla S - \kappa_r\varrho C_p\nabla T
\end{equation}\end{linenomath}
with the entropy diffusion coefficient $\kappa$ and the radiative diffusivity $\kappa_r$.
The specific entropy $S$ is defined as
\begin{linenomath}\begin{equation}
S=C_p {\rm ln}\left(P^{1/\gamma}/\varrho\right)
\end{equation}\end{linenomath}
with the specific heat at constant pressure $C_p$.
The extra terms, $\varrho\vec{g}$ and $-2\varrho\vec{\Omega_o}\times\vec{U}$, are for the gravity and the Coriolis force, respectively.

The impenetrable plasma flow, $V_r=\vec{U}\cdot\hat{r}=0$ ($\hat{r}$ is the unit vector of the heliocentric position vector $\vec{r}$, $\hat{r}:=\vec{r}/|\vec{r}|$, or the outward-pointing normal unit vector to the boundary surface), is applied to the top and bottom boundary surfaces.
In addition, the stress-free condition is achieved by
resetting the shear stress components associated with the normal direction to the boundary sphere
($\tau_{r,\theta}$, $\tau_{r,\phi}$, $\tau_{\theta,r}$, and $\tau_{\phi,r}=0$ when writing the terms in the spherical coordinate system) to zero.
On the bottom boundary surface, the heat flux ${\bf q}$
is fixed at $L_\sun/(4\pi R_{\rm bot}^2)\hat{r}$.

In the CHORUS++ code, the velocity vector $\vec{U}$ is defined in the Cartesian system,
$\vec{U}=(u,v,w)$ or 
$\vec{U}=(V_x,V_y,V_z)$,  
and the numerical fluxes normal to the cell faces are calculated through the Bassi-Rebay I scheme \citep[][]{Bassi_1997}.
The governing equations in the physical domain are transformed into local coordinates of a computational cell using the transfinite mapping approach.
The transfinite mapping approach is a major improvement over the original CHORUS code \citep[][]{Wang15} because it can represent the curves of the edges and faces of each physical element within the machine precision \citep[][]{Chen23}.
In the presented study, we set the grid size in the radial direction at about 0.001563 $R_\sun$ and the heliocentric angles of the numerical cells at 0.4 $\sim$ 0.5 degrees.

\subsection{Model parameters, initial values, and simulation setup}
We used the same method as in \citet{Chen23} to set up the initial values, except that the stellar luminosity, rotation rate, the radius of the upper boundary sphere surface, and the initial plasma density at the upper boundary surface are altered.
As in \citet{Chen23}, the initial radial profiles of the plasma quantities are generated in accordance with the so-called laminar benchmark \citep[][]{Jones11}.
In conjunction with the change of the radius of the top boundary surface ($r=R_{\rm top}$), we need to adjust the plasma density there.
The parameters set in the presented study are tabulated in Table \ref{tblparam}.
For reference, the parameters used in the earlier work using the CHORUS++ code \citep[][]{Chen23} and some of the actual solar properties are also given in this table.

\begin{table}[h]
\caption{%
Parameters, constants, and representative numbers for (a) the actual Sun, (b) the presented case, 
and (c) 
the numbers used in \citet{Chen23} (and/or \citet{Wang15}).
The constants are set as follows:
Gravitational constant $G=6.67\cdot 10^{-8}$ (${\rm cm}^3 {\rm g}^{-1}{\rm s}^{-2}$),
the solar mass $M_\sun=1.98891\cdot 10^{33}$ (g),
the specific heat ratio $\gamma= 5/3$,
the gas constant ${\cal R}=1.4\cdot 10^8$ (erg ${\rm g}^{-1} {\rm K}^{-1}$,
the specific heat at constant pressure $C_P=3.5\cdot 10^8$ (erg ${\rm g}^{-1} {\rm K}^{-1}$),
the solar luminosity $L_\sun=3.846\cdot 10^{33}$ (erg\,s$^{-1}$), 
and 
the solar sidereal rotation rate $\Omega_\sun=3.0\cdot 10^{-6}$ (rad\,s$^{-1}$).
The parameter $D$ is the thickness of the convection zone ($D=R_{\rm top}-R_{\rm bot}$).
}%
\centering
\begin{tabular}{l|r|r|r}
\hline
   & (a) the Sun & (b) presented case & (c) \citet{Chen23} \\
\hline
 fixed time step, $\Delta t [s]$ &  & 2.5 &  4 $\sim $ 20  \\
 Lower bound., $R_{\rm bot}$ in $R_\sun$ & $\sim$0.70 & 0.70 & 0.70 \\
 Upper bound., $R_{\rm top}$ in $R_\sun$ & 1 & 0.99 & 0.95 \\
 Density at $r=R_{\rm bot}$, $\varrho_{\rm top}$ $[g\cdot cm^{-3}]$ & &
                             $2.1\cdot 10^{-1}$ &
                             $2.1\cdot 10^{-1}$ \\
 Density at $r=R_{\rm top}$, $\varrho_{\rm top}$ $[g\cdot cm^{-3}]$ & &
                             $2.1\cdot 10^{-4}$ & $1.05\cdot 10^{-2}$ \\
 Den. scale, $N_p=\log_e(\varrho_{\rm bot}/\varrho_{\rm top})$ &
   $\sim$ 16 &
   6.908 ($\sim\log_e 1000$)&
   3.0 ($\sim\log_e 20$)\\
 Luminosity $L_o$ &
   $1\times L_\sun$ &
   $1.0\times L_\sun$ &
   $10^3\times L_\sun$ \\
 Rotation rate, $\Omega_o$ &
   $\sim 1 \times\Omega_\sun$&
   $1.0 \times\Omega_\sun$ &
   $27.0\times\Omega_\sun$ \\
 kinematic viscosity $\nu$ $[cm^2 s^{-1}]$ & & 
   $6.0\cdot 10^{12}$ &
   $6.0\cdot 10^{13}$ \\
 entropy diffusion coef. $\kappa_s$ $[cm^2 s^{-1}]$ & &
   $6.0\cdot 10^{13}$ &
   $6.0\cdot 10^{13}$ \\
 \hline
 \hline
 $U'_{\rm rms}$ [cm$\cdot$s$^{-1}$] & & 
   $1.51\cdot 10^4$ &
   $4.04\cdot 10^4$ \\
 $\Delta S$ [erg/g/K] & & 
   $1.249\cdot 10^6$ &
   $7.798\cdot 10^5$ \\
 \hline
 \hline
 Rayleigh num. $GM_\sun D \Delta S/(\nu\kappa_s C_p)$  &
   $10^{20}$ &
   $2.670\cdot 10^7 $ &
   $1.429\cdot 10^6$ \\
 Reynolds num. $U'_{\rm rms} D / \nu$ &
   $10^{12}$ &
   $51.06 $ &
   $11.72 $ \\
 Ekman num. $\nu/(\Omega_o D^2)$  &
   $10^{-14}$&
   $4.853\cdot 10^{-3} $ &
   $2.447\cdot 10^{-3} $\\
 Taylor num. $4\Omega_o^2 D^4/\nu^2$ &
   $10^{19\sim 27}$ &
   $1.698\cdot 10^5$ &
   $6.682\cdot 10^5$ \\
 Prandtl num. $\nu/\kappa_s$ &
   $10^{-6\sim -4}$
   & $0.1$
   & $1 $\\
 Rossby num. $U'_{\rm rms}/(2\Omega_o D)$ &
   $10^{-1\sim 0}$ &
   $1.239\cdot 10^{-1}$ &
   $1.433\cdot 10^{-2}$ \\
\hline
\end{tabular}
\label{tblparam}
\end{table}

We set a lower value for the kinematic viscosity ($\nu=6\cdot 10^{12}$) than that in \citet{Chen23} ($\nu=6\cdot 10^{13}$) to make the simulated system more closely match the actual Sun.
Nonetheless, the scale parameters, such as the Rayleigh number and the Reynolds number, are different from those for the solar numbers by several to ten orders of magnitude.
The chosen kinematic viscosity is the smallest value with which the simulation could run without numerical instability.
Finer mesh assignments will allow us to simulate with a smaller value of the coefficient for the second-order derivatives; however, the use of such realistic viscosity or diffusivity would be left for future work.
Given the current computational resources available, we set only the total luminosity and the rotation rate to those of the actual Sun.

The initial values are determined in the same manner as described in \citet{Chen23}.
In brief, the initial velocity is set to zero, and the plasma density, gas pressure, and temperature are determined such that the polytropic relation, the heat flux conservation, and the balance between the radial pressure gradient and gravity will be satisfied.
The calculation of the initial values requires the thermal heat conduction rate ($\kappa_r$) as a function of the radius and the density scale ($N_p$; the ratio of the mass density at the upper and lower boundary surfaces).
We used the same values of the thermal heat conduction rate as in \citet{Chen23} but a different value of the density scale (because of the difference in the radius of the top boundary surface, $R_{\rm top}$).

\begin{figure}
\epsscale{0.8}
\plotone{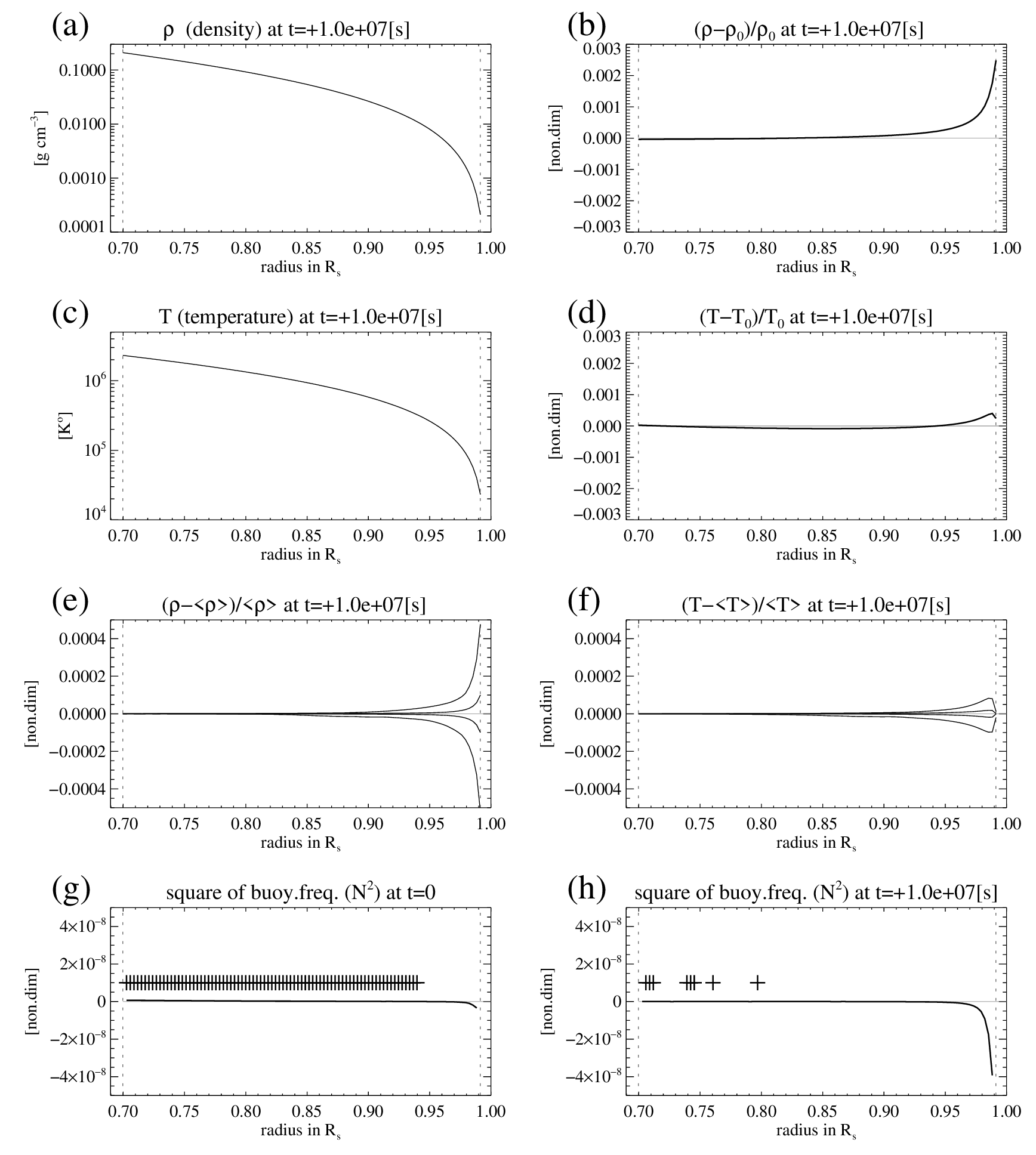}
\caption{%
Radial profile of the simulated plasma quantities in the solar convective shell at $t=1\cdot 10^{7} s$:
(a) mass density $\varrho$ and (c) temperature ($T$).
In panels (b) and (d), the deviations of the simulated variable from the initial values,
$\varrho_0$ and $T_0$,
are given as $(\varrho/\varrho_0-1)$ and $(T/T_0-1)$, respectively.
The deviations from the initial radial profile are less than $3\cdot 10^{-5}$ at $r < 0.9R_\sun$ and about $2\cdot10^{-3}$ near the top boundary.
The contrast of the plasma density is tiny as
$|\varrho/\left<\varrho\right>-1| < 3\cdot 10^{-4}$,
where $\left<\varrho\right>$ is the averaged density over the spherical layer.
Panels (e) and (f) show the relative deviation of the mass density and temperature from the averages,
$(\varrho-\left<\varrho\right>)/\left<\varrho\right>$ and
$(T_p    -\left<T_p    \right>)/\left<T_p    \right>$, respectively, with four lines representing the maximum, +1 standard deviation, -1 standard deviation, and the minimum values at each radius.
In the bottom row, the radial profile of the square of buoyancy frequency ($N^2$) of the initial value
and the buoyancy frequency averaged horizontally (latitudinally and longitudinally) are given in panels (e) and (f), respectively.
In these two plots, cross (+) marks are placed where $N^2 > 0$.
}%
\label{fig01}
\end{figure}
\begin{figure}
\epsscale{0.5}
\plotone{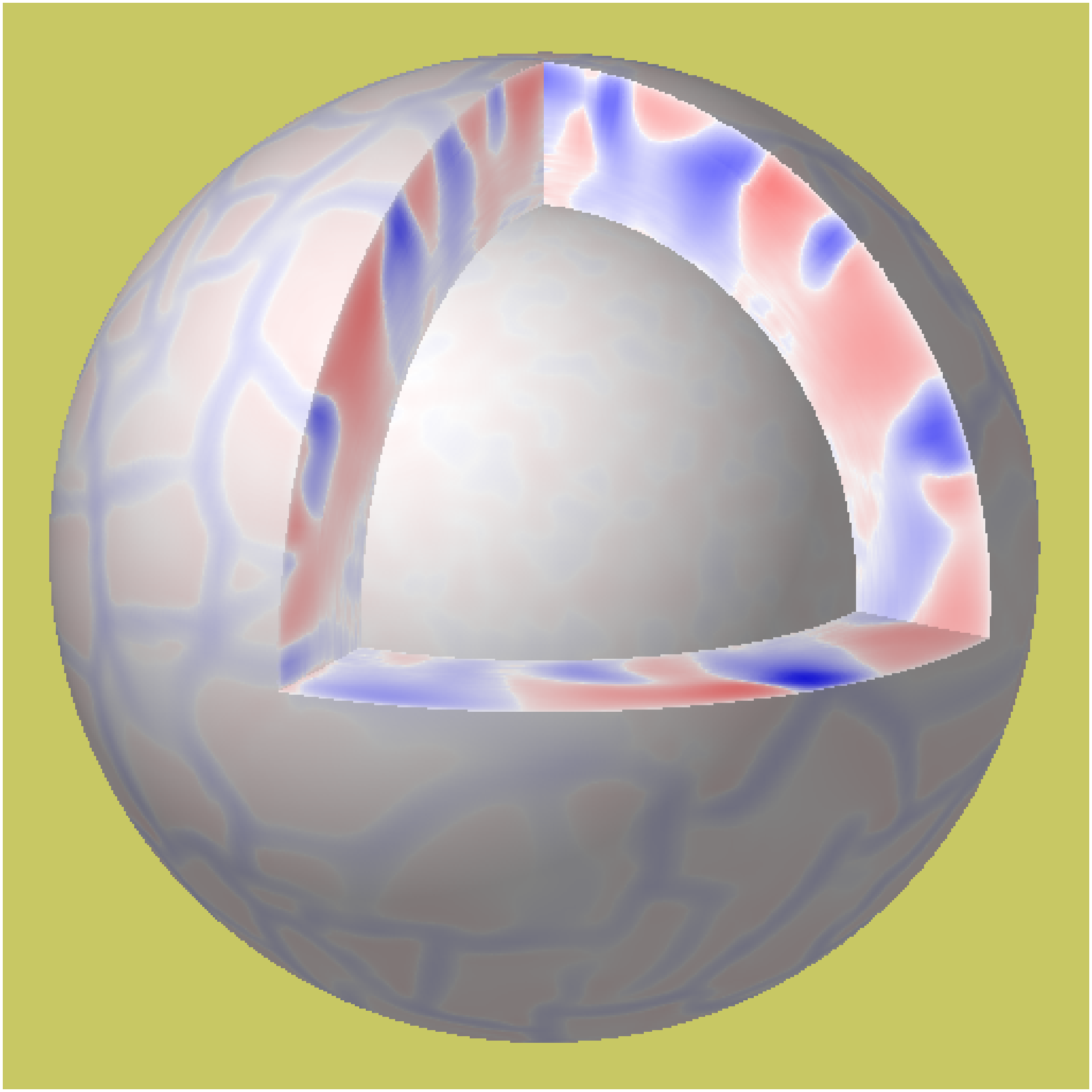}
\caption{%
The simulated radial component of the plasma flow ($V_r$) at  $t=10^{7} s$.
An octant section has been removed to display the meridional and equatorial cross-sections.
The red (blue) colors represent the positive (negative) values.
}%
\label{fig02}
\end{figure}

The initial state is nearly stable except in the near-surface regions at $r \sim R_{\rm top}$ where the plasma structure is unstable to perturbations or fluctuations.
Figure \ref{fig01}(g) shows the radial profile of the square of the buoyancy frequency at the initial value,
\begin{linenomath}\begin{equation}
N^2
=g(r)\left[
{1\over\gamma}{1\over P_g}{\partial P_g\over\partial r}-{1\over\varrho}{\partial\varrho\over\partial r}
\right],
\end{equation}\end{linenomath}
where the positive (negative) $N^2$ indicates that the region is stable (unstable) against perturbations.
The initial plasma distribution is unstable at $r > 0.94 R_\sun$.
The instability gradually grows, and the convective flows ultimately prevail throughout the simulated convection zone.

Figure \ref{fig02} demonstrates the radial component of the plasma velocity ($V_r$) simulated at $t=1\cdot 10^7s$.
The convective flows are seen in the simulated convection zone, as well as the convective cell structures on the top boundary surface.
In Figure \ref{fig01}(h), the square of the buoyancy frequency $N^2$ is negative nearly all over the simulation volume at the selected moment.

Panels (a) and (c) of Figure \ref{fig01}, respectively, show the radial profile of the mass density and temperature averaged over each shell layer at $t=1\cdot 10^7$s.
Panels (b) and (d), respectively, show the ratios of the simulated plasma density and temperature to those of the initial values.
A notable difference is found in the density ratio near the top boundary surface, as shown in panel (b); still, the deviation is 0.25\% in the largest.
In panels (e) and (f) of Figure \ref{fig01}, the deviations of the plasma density and temperature from the average profile are small, too; the largest relative deviation of density and temperature,
$(\varrho-\left<\varrho\right>)/\left<\varrho\right>$ and
$(T_p    -\left<T_p    \right>)/\left<T_p    \right>$, are estimated at
$\sim 5\cdot 10^{-4}$ and $\sim 1\cdot 10^{-4}$, respectively, except for the regions close to the 0.99-$R_\sun$ top boundary surface.

\section{Results}\label{sctresults}
The kinetic energy density ($\varrho V^2/2$) is a good proxy to represent the entire simulated convective system.
In \citet[][]{Chen23}, the kinetic energy density of the simulated stellar convection zone exhibits exponential growth and a brief overshoot/oversaturation, then reaches a plateau profile that corresponds to the statistically stable saturated convection state.

\begin{figure}
\epsscale{0.75}
\plotone{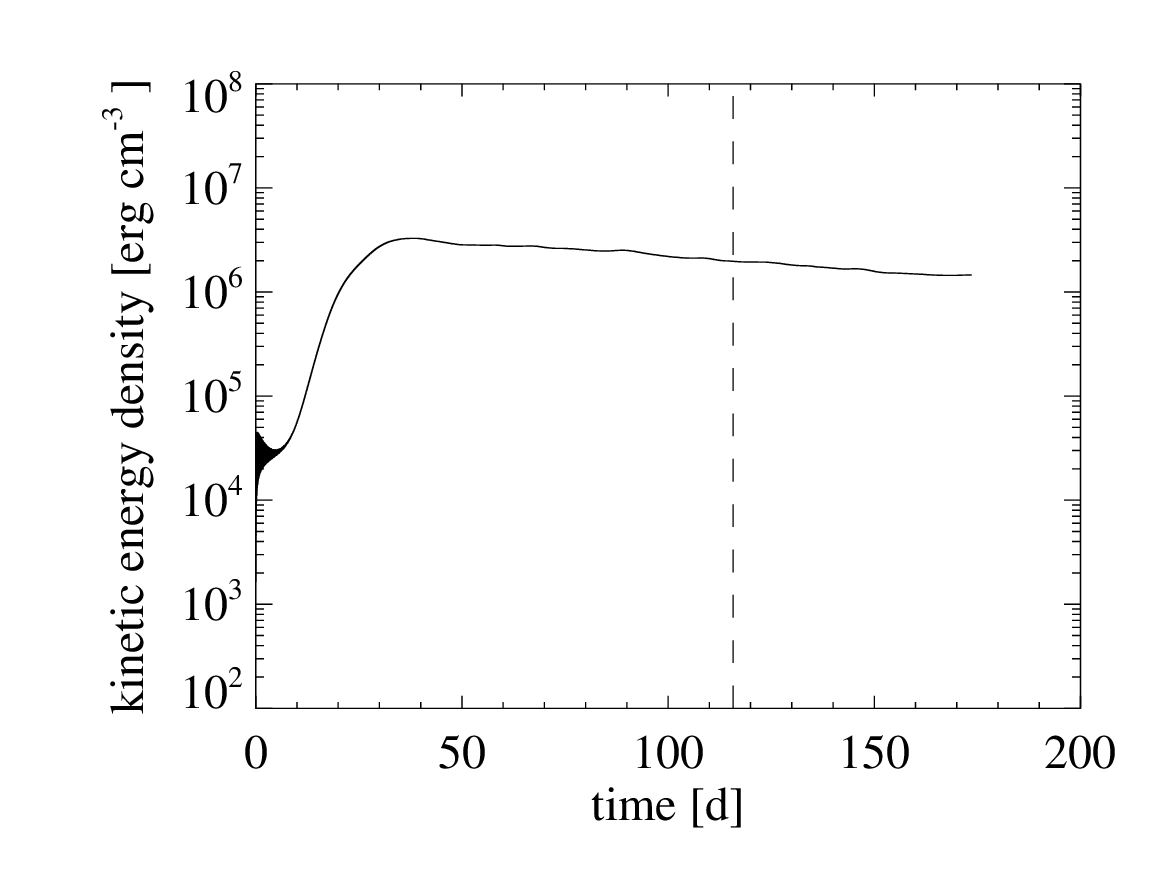} 
\caption{%
Temporal profile of the average kinetic energy density. The vertical dashed line indicates the moment ($t \simeq \cdot 10^7$\,s for which most figures in this article are made using the simulation data.
}%
\label{fig03}
\end{figure}

Figure \ref{fig03} shows the temporal profile of the kinetic energy density.
With the actual solar parameters, the temporal profile of the kinetic energy density is different from that
in \citet{Chen23}.
On the logarithmic scale plot, the exponential growth of the energy density is found as a (near) linear line, roughly at $10 \le t \le 20$ days.
However, the overshooting and the following plateau profile \citep[like those in Figure 7 of ][]{Chen23} are not clearly seen.
The current simulation run has been performed only for $t \le 1.5\cdot10^7~{\rm s} \sim 170~{\rm d}$, but it is unclear whether or not the kinetic density will become more constant in an extended simulation run.
Because our visual inspections did not find distinct differences in the figures for $t = 1\cdot10^7 s$ and $t = 1.5\cdot10^7 s$, we chose the simulated data at $t=10^7\,{\rm s} \sim 115 \,{\rm d}$ as a moderately mature state of the convection properties.

\subsection{Large-scale Convection Cells}
\begin{figure}
\epsscale{1.0}
\plotone{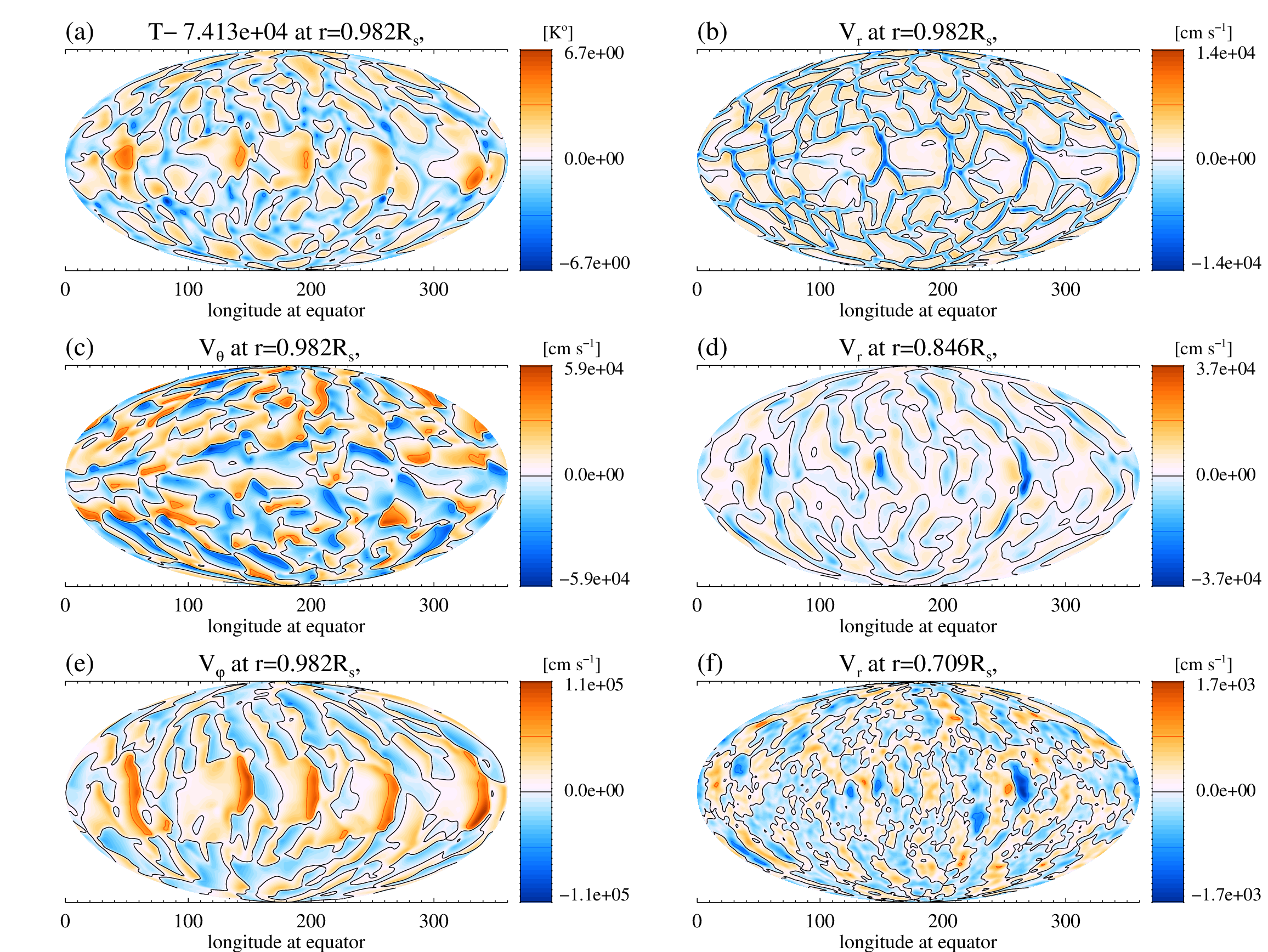}
\caption{%
The plasma quantities of the simulated solar convection zone at $t=10^{7}$\,s in the Mollweide projection mapping.
(a) The plasma temperature subtracted by the average ($7.4132\cdot 10^4$\,K).
(b), (c), and (e) The radial, latitudinal, and longitudinal components of the plasma flow ($V_r$, $V_\theta$, and $V_\phi$),
at $r=0.982\,R_\sun$.
Panels (b), (d) and (f) in the right column compare the radial component at three depths,
near the top boundary, at the middle depth of the simulated convection zone (at $r=0.846\,R_\sun$), and near the bottom boundary surface (at $r=0.709\,R_\sun$).
The colors are truncated at the largest absolute value so that the red (for positive values) and blue (for negative values) are assigned evenly across the zero values.
}%
\label{fig04}
\end{figure}

Figure \ref{fig04}(a) shows the deviation of temperature from the averaged value at a selected near-surface depth (at $r=0.982\,R_\sun$ and  $t=1.0\cdot 10^7\,{\rm s}$).

As shown in Figure \ref{fig01}(b,d), the deviations of the simulated plasma variables from the initial values are relatively small in the radial direction.
The temperature, averaged over the selected sphere, was $7.4132\cdot 10^4\, {\rm K}$, approximately 8 degrees higher than the initial value, and the largest deviation from the averaged state is about 6.7 degrees.
The lower temperature structures are found at the boundaries of the convection cells that can be identified as negative velocity $V_r$ in Figure~\ref{fig04}(b).
The temperature and plasma motions near the surface are found to be rather uneven within convection cell structures.
For example, the region of relatively high temperature (panel a) and large prograde plasma motion ($V_\phi$ in panel e) is found in the western (right) part of the convection cells near the equator (panel b).
The latitudinal motion ($V_\theta$ in panel c) exhibits a weak but similar tendency to $V_\phi$; relatively large poleward flows are found near the boundary of each convective cell close to the poles in each hemisphere.
In the right column in Figure~\ref{fig04}(b,d,f), the distribution of $V_r$ at different depths is presented.
The size and shape of the convection cells differ substantially at different depths, while several strong downward flows are found at the same locations over a wide range of depths.

\begin{figure}
\epsscale{1.0}
\plotone{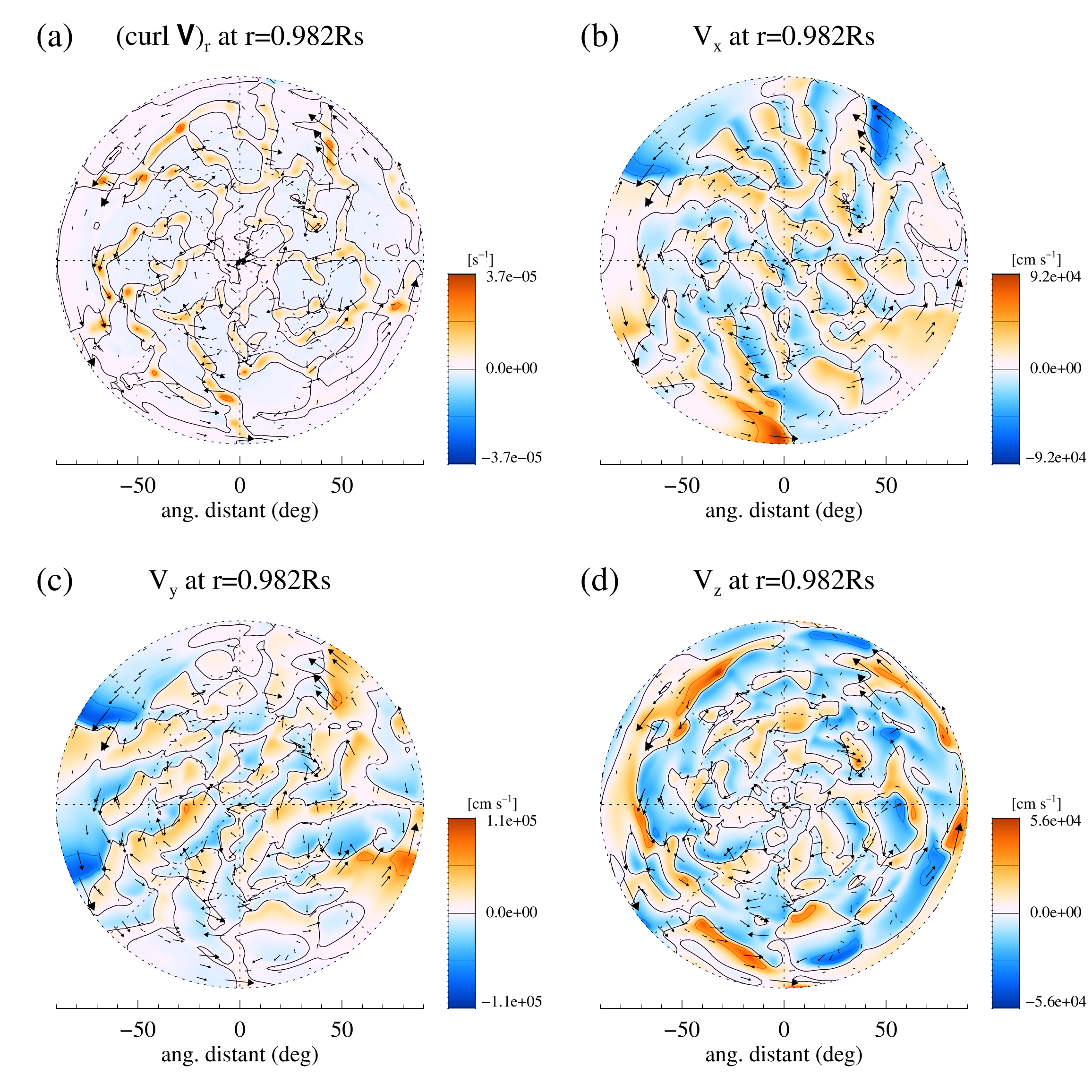}
\caption{%
(a) The radial component of the curl of plasma velocity (or the curl of the horizontal flows).
(b) --- (d) The $x$, $y$, and $z$ components of the simulated plasma flows around the north pole.
}%
\label{fig05}
\end{figure}

\subsection{Polar regions}
Figure \ref{fig05} offers the north pole view of the simulated plasma flows near the top boundary surface (at $r=0.982\,R_\sun$).
Because the CHORUS++ code uses an unstructured grid system, issues due to the geometric singularity, like those in the case of the spherical coordinate formulation, are not present, and the plasma flows crossing the solar rotation axis (or $z$-axis) are treated naturally.
The snapshot shown (at $t=1\cdot 10^7$\,s) does not show clear polar vortex cell structures \citep[e.g.,][]{Nagashima11,Dikpati24}.
By tuning our future simulation settings, we plan to investigate in more detail the formation of plasma-flow structures in the high-latitude and polar regions, which are believed to play a major role in solar-cycle activities.

\begin{figure}
\epsscale{1.0}
\plotone{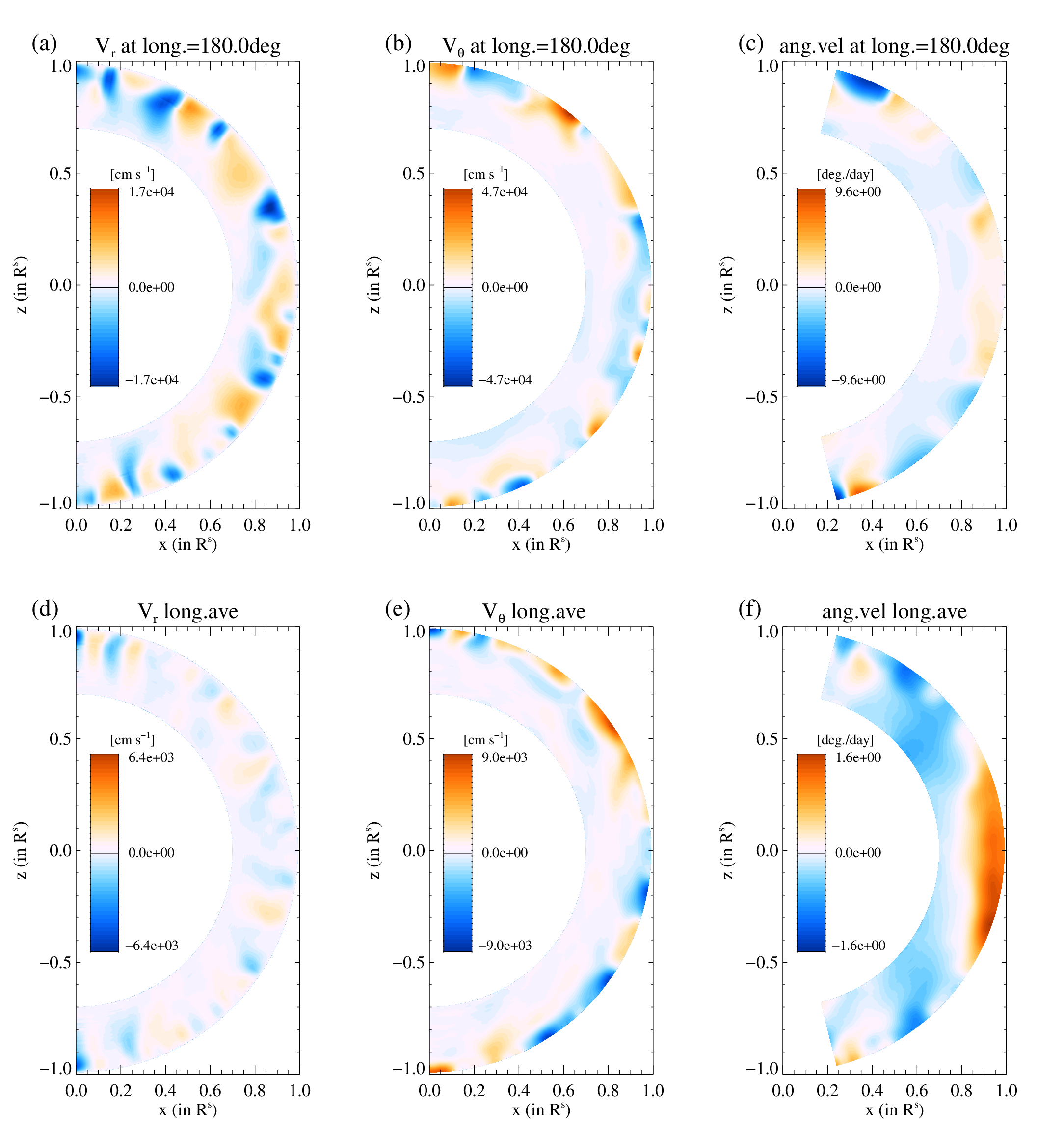}
\caption{%
The simulated plasma flows
on the meridional plane
at $t=10^{7} s$.
From left, the radial component of the plasma flow ($V_r$), 
the latitudinal component ($V_\theta$), and the rotation rate relative to the reference frame ($\omega:=V_\phi/(r\sin\theta)$) are shown.
In the top row (a) -- (c), the values are sampled at a selected longitude ($\phi=$180 degrees).
In the bottom row (d) -- (f), the average over the longitude are shown.
}%
\label{fig06}
\end{figure}

\subsection{Meridional Circulation and Rotation}
Figure \ref{fig06} shows the radial and latitudinal components of the plasma flow vector ($V_r$ and $V_\theta$, respectively) and the angular velocity with respect to the simulation coordinate frame, $\omega:=V_\phi/r\sin\theta$.
In the upper row, the flow components at a selected longitude ($\phi=180^\circ$) are shown.
In panel (a), the downward flows (negative $V_r$) starting from near the solar surface are overall bent equatorward before reaching deeper layers.
A similar trend is found in the longitudinally averaged values shown in panel (d).

In Figure~\ref{fig06}(e), the averaged latitudinal component ($V_\theta$) overall exhibits the equatorward motions near the solar surface.
Counter (poleward) flows are found at a relatively shallow depth $r\sim 0.85R_\sun$ in the northern hemisphere.
The values of the latitudinal component $V_\theta$ at deeper layers are small. Hence, the patterns of the equatorward bents of $V_r$ seen in panel (a) have been caused by the poleward motion of the near-surface sources rather than the global (averaged) advection motion.

In Figure~\ref{fig06}(f), the distributions of the averaged angular rotation rate are more similar to those inferred from observations: faster (prograde) equator rotation near the surface and slower (retrograde) high-latitude motions. 
The difference between the rotation rates at the solar equator and middle latitude region ($\pm 30^\circ$ from the equator) is about 1.5 degrees per day or 40 degrees per one solar rotation, which is much larger than that of the actual Sun.

\begin{figure}
\epsscale{1.0}
\plotone{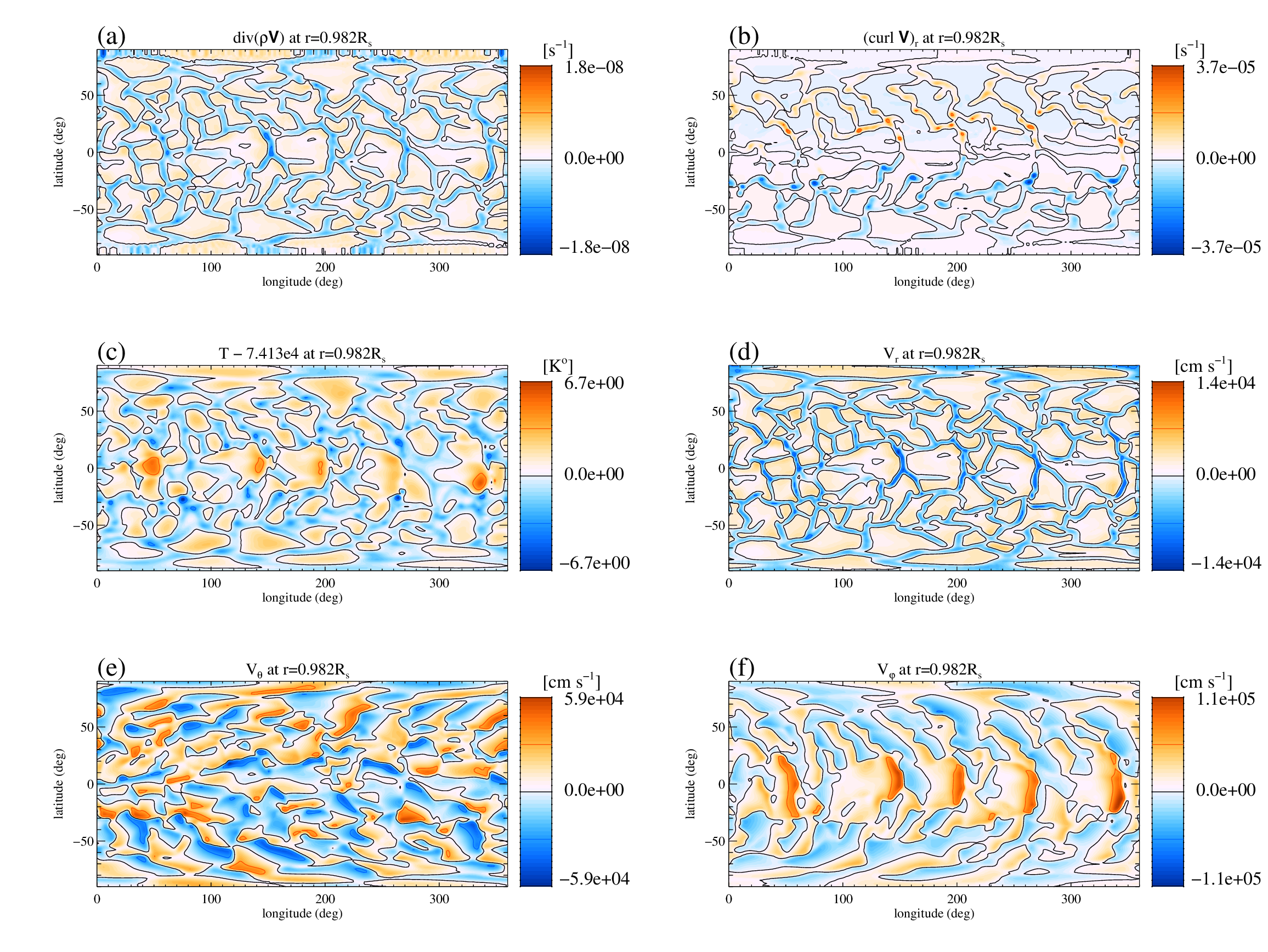}
\caption{%
Latitude-longitude plots of 
(a) the divergence of mass flux ($\nabla\cdot (\varrho \vec{V})$),
(b) the radial component of the curl of plasma velocity ($\hat{r}\cdot(\nabla\times\vec{V})$),
(c) the temperature subtracted by the average ($7.4132\cdot 10^4$\,K),
and
(d) -- (f) the radial, latitudinal, and longitudinal components of the plasma velocity ($V_r$,$V_\theta$,$V_\phi$)
at $r=0.982\,R_\sun$ and $t=10^{7}$\,s.
The distributions of the spatial derivatives shown in panels (a) and (b) appear correlated with that of $V_r$ in panel (d).
The colors are truncated at the largest absolute value so that the red (for positive values) and blue (for negative values) are assigned evenly across the zero values.
}%
\label{fig07}
\end{figure}

\subsection{Divergence and vorticity of plasma flows}\label{sctdivrhovel}
The present simulation solves the fully compressible HD equations; hence, no assumptions are given to the divergence of the plasma flow, $\nabla\cdot\vec{V}$, or the mass flux, $\nabla\cdot(\varrho\vec{V})$.
Figure \ref{fig07} shows the latitude-longitude distribution of plasma flows near the top boundary surface.
In panels (a) and (b), the divergence of the mass flux and the radial component of the vorticity of the plasma flow vector, $\nabla_h\times\vec{V}_h$, are shown, respectively.
The convergence (negative divergence) of the mass flux is found at
the region of the downward flow (negative $V_r$) and lower temperature,
or the boundary of the large-scale convection cells.

In Figure~\ref{fig07}(b), the distribution of the radial component of the vorticity $(\nabla\times\vec{V})_r$ exhibits interesting features.
In the northern hemisphere, for example, the positive (counter-clockwise vortices viewed from outside) values are concentrated along the boundaries of the convection cells (where $V_r<0$), while rather negative values are found evenly within the cells over the hemisphere.
In the southern hemisphere, the same tendency, except for the signs of the vector quantities, is found.
In the northern (southern) hemisphere of the examined state, the Coriolis force initiates the clockwise (counterclockwise) motions of the divergent flows of the convection cells, and the positive (negative) vorticity is formed where two negative(positive) vorticity regions contact each other.
In panels (c) and (f) of Fig.\ref{fig06}, we obtained a solar-like differential rotation pattern near the solar surface at the examined system state.
This is because the positive vorticity is dominant over the negative one.
It is necessary to conduct the simulation for a longer time to assess how long such unbalanced vorticity can persist and whether or not the vorticity imbalance can be found in a mature stage of global convection.

\begin{figure}
\epsscale{1.0}
\plotone{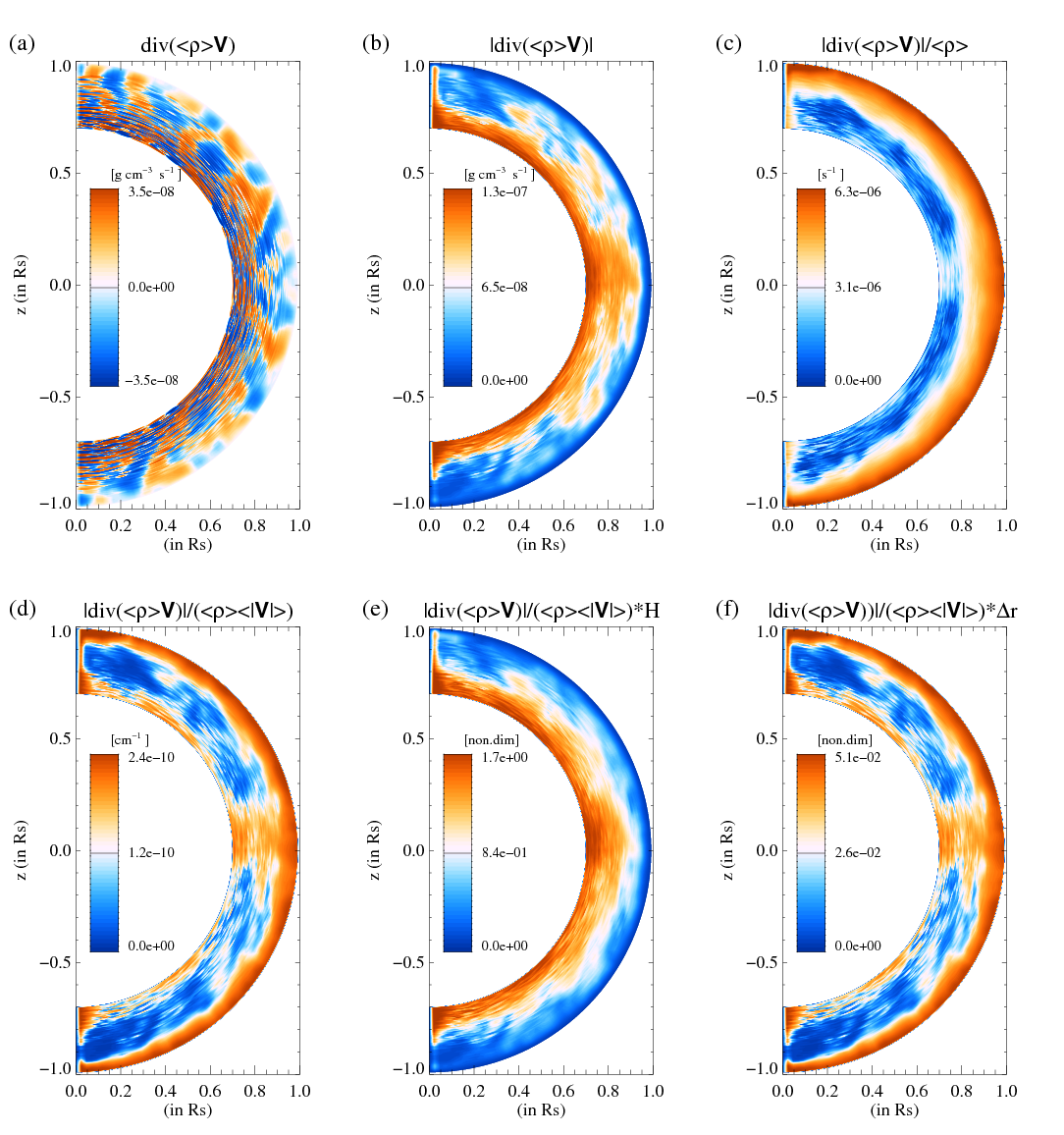}
\caption{%
The longitudinal averages of
(a) the divergence of mass flux with the average of mass density at each height ($\left<\varrho\right>$),
$ \nabla\cdot(\left<\varrho\right>\vec{V})$,
(b) the absolute value of the divergence,
$|\nabla\cdot(\left<\varrho\right>\vec{V})|$,
(c) the absolute divergence normalized with the mass density average,
$|\nabla\cdot(\left<\varrho\right>\vec{V})|/ \left<\varrho\right>$,
(d) the absolute divergence normalized with the average of the absolute mass flux,
$|\nabla\cdot(\left<\varrho\right>\vec{V})|/(\left<\varrho\right> \left<|\vec{V}|\right>)$,
and (e) and (f) the same quantity as in (d) but multiplied by the scale height ($H$), and the radial grid size ($\Delta r$), respectively.
}%
\label{fig08}
\end{figure}

To assess
the differences between the presented model and anelastic ones, we calculate the divergence of $(\left<\varrho\right>\vec{V})$, where $\left<\varrho\right>$ is the density averaged over each sphere of the same heliocentric distance.
Figure \ref{fig08}(a) shows the meridional view of the divergence of the mass flux averaged over the longitude.
In panel (b), the average of the absolute divergence, $|\nabla\cdot(\left<\varrho\right>\vec{V})|$, is shown.
In Figure~\ref{fig08}(c), the average of the absolute values is normalized with the average of the mass density, 
$|\nabla\cdot(\left<\varrho\right>\vec{V})|/\left<\varrho\right>$, 
so that the normalized value will be in the physical unit of ${\rm s}^{-1}$, and the inverse corresponds to the time scale for the divergence of the mass flux to substantially alter the local density.
In this case, the larger values are found near the outer boundary surface, meaning that the relative variations of the plasma density are greater in the outer part of the convection zone than near the bottom of the convection zone, which is rather anticipated.
The smallest value of the time scale (or the inverse of the largest value of the normalized divergence) is found near the top boundary surface and estimated at about $1.6\cdot 10^5$\,seconds or about 1.8 days.
The time scale for the deeper layers is estimated to be of an order of $10^4\,{\rm days}$.
These time-scale estimates imply that simulations with a divergence-free mass flux can properly handle the simulated system at least for this limit in a spherical convective shell, which does not include near-surface and surface layers. 
Of course, the size and position of the convection cell change quickly, so the practical limit must be much longer.

Figure~\ref{fig08}(d) offers the distribution of $|\nabla\cdot(\left<\varrho\right>\vec{V})|/(\left<\varrho\right>\left<|\vec{V}|\right>)$.
This quantity has a physical unit of $\rm{cm}^{-1}$, and the inverse represents the spatial scale of the contribution of the divergence of the mass flux.
The largest values are found near the top boundary, and the maxima are about $2.5\cdot 10^{-10}\, {\rm cm}^{-1}$, corresponding to the spatial scale of about $4\cdot 10^9 {\rm cm}$, comparable to the thickness of the simulated convection zone.
Near the bottom boundary surface, the spatial scale is estimated to be larger.
The estimated spatial scales suggest that the discrepancy between the incompressible and compressible models is not significant for the global scale of the solar convection simulations with the actual solar parameters.

Figure~\ref{fig08}(e) shows the quantity shown in panel (d) multiplied by the local scale height, $H=(d\ln\varrho/dr)^{-1}$, to assess the degree of discrepancy in small volumes.
The regions with values larger than 1.0 are found in the inner part of the simulated convection zone.
This implies that compressibility can be one of the primary factors in determining the high-density slow-flow regions in the long term.
Near the top boundary surface, the values are much smaller, indicating that the divergence of mass flux is not sufficiently large to substantially alter the background radial profile of the plasma density. 

Lastly, Figure~\ref{fig08}(f) shows similar quantities to those shown in panel (e), except that the size of the numerical cells in the radial direction is used to multiply the values in panel (d).
This dimensionless value is always less than 0.063, meaning that the cell size is sufficiently small to handle the divergence of the mass flux and, hence, the temporal variation of the mass density properly.

\begin{figure}
\epsscale{1.0}
\plotone{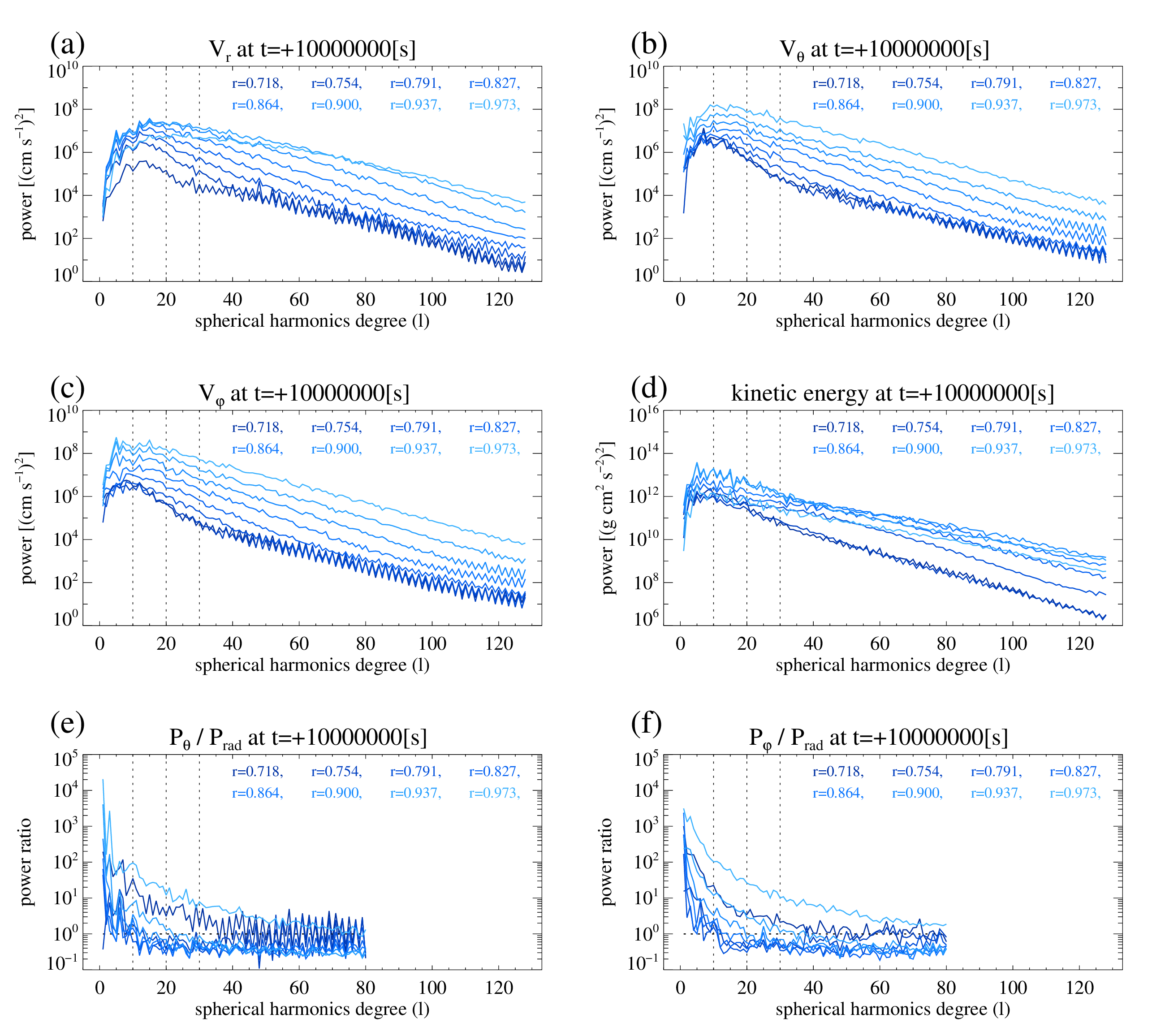}
\caption{%
(a) -- (d) Spherical harmonics spectral of three components of the plasma flow velocity ($V_r$, $V_\theta$, and $V_\phi$) and the kinetic energy density ($\varrho V^2/2$) at various depths.
(e) The ratio of the power of $V_\theta$ (shown in panel (b)) to that of $V_r$ (shown in panel (a)).
(f) The ratio of the power of $V_\phi$ (shown in panel (c)) to that of $V_r$.
In panels (e) and (f), the plot lines are truncated at $l=80$.
The sampling depths are indicated with color brightness; the darker (lighter) blue for deeper (shallower) layers.
}%
\label{fig09}
\end{figure}

\subsection{Spatial scale in the latitudinal and longitudinal directions}
Panels (a) -- (d) of Figure~\ref{fig09} show the spherical harmonic spectral powers of the three components of the plasma velocity ($V_r$, $V_\theta$, and $V_\phi$) and the kinetic energy density ($\varrho V^2/2$).
Overall, the peaks are found around the harmonic degree ($l$) from 5 to 15 at any depths, and the peak values of these four parameters are larger (smaller) at the outer (inner) parts.
The simulation grid size in the horizontal (perpendicular to the radial direction) is about 1 degree in the heliocentric angle, and a typical horizontal structure is captured by approximately 30 subgrid cells.
The powers for the longitudinal component $V_\phi$ and the kinetic energy ($\varrho V^2/2$) have peaks around $l\sim 5$, while the other two ($V_r$ and $V_\theta$) have peaks around $l\sim 10$.
The powers of the plasma flow velocity become smaller as the sampling depth becomes deeper, while this tendency is not clear in the kinetic energy density.

In panels (e) and (f), the ratios of the powers of $V_\theta$ to $V_r$ and $V_\theta$ to $V_\phi$ are shown.
If the three components are evenly driven then the ratio would be close to 1.
As seen in these two panels, the power ratios are seldom equal or close to 1 ($10^0$); hence, the anisotropy in the convection zone \citep[e.g.,][]{Guerrero22} is significant.
In the presented simulation, the radial component of plasma flow is set to zero ($V_r=0$) at the bottom and top boundary surfaces; hence, these ratios for depths closer to the boundaries tend to be larger.

At the depths away from the boundary surfaces, overall, the ratio is greater than 1 for $l < 10 \sim 20$, implying that the horizontal flows dominate over the radial flows in the large scale over the wide range of depths.
For larger harmonic degrees (i.e., $l>20$), the power ratios are rather constant at $10^{-0.3}\sim10^{-0.4}$ or 0.4 $\sim$ 0.5 over the different depths; hence, the square of the radial component $V_r^2$ and that of horizontal components $V^2_h=V_\theta^2+V_\phi^2$ are nearly balanced, indicating that the anisotropy of the plasma flows is small in the small spatial scale over a wide range of radius.

\begin{figure}
\epsscale{1.0}
\plotone{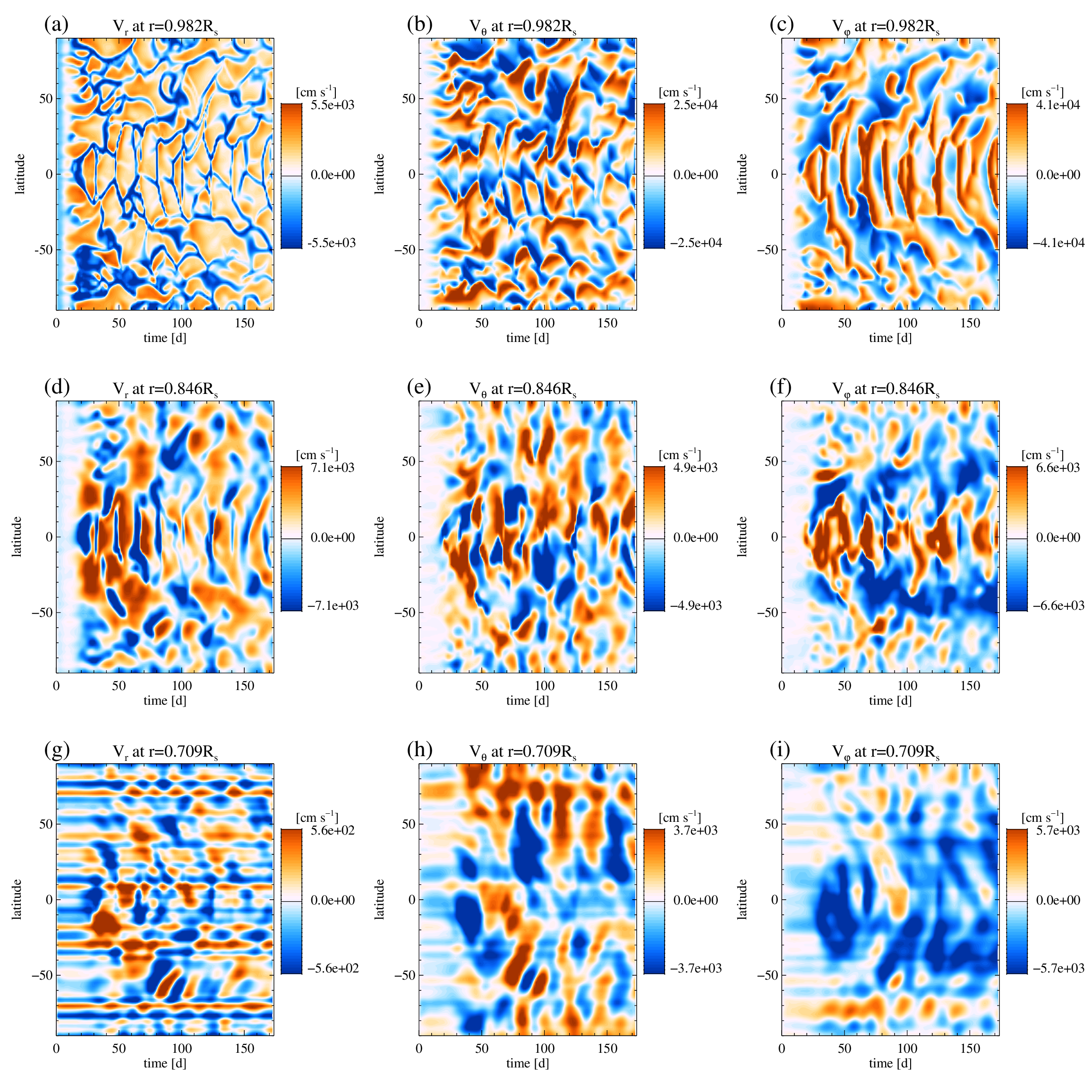}
\caption{%
Latitude-time diagram of the plasma velocity at a selected longitude (180 degrees).
In the top row (a) -- (c), the three components of the plasma velocity
($V_r$, $V_\theta$, $V_\phi$) sampled at $r=0.92R_\sun$ (near the top boundary surface) are shown.
In the middle (d) -- (f) and bottom row (g) -- (i), the same quantities but sampled at $r=0.846R_\sun$ (the middle depth of the simulated domain) and at $r=0.709R_\sun$ (near the bottom boundary surface) are shown, respectively.
The time runs from left to right.
In panels (a) and (c), poleward motions of the patterns are found clearly.
The colors were truncated at the 95th percentile of the absolute value of each variable.
}%
\label{fig10}
\end{figure}

\begin{figure}
\epsscale{1.0}
\plotone{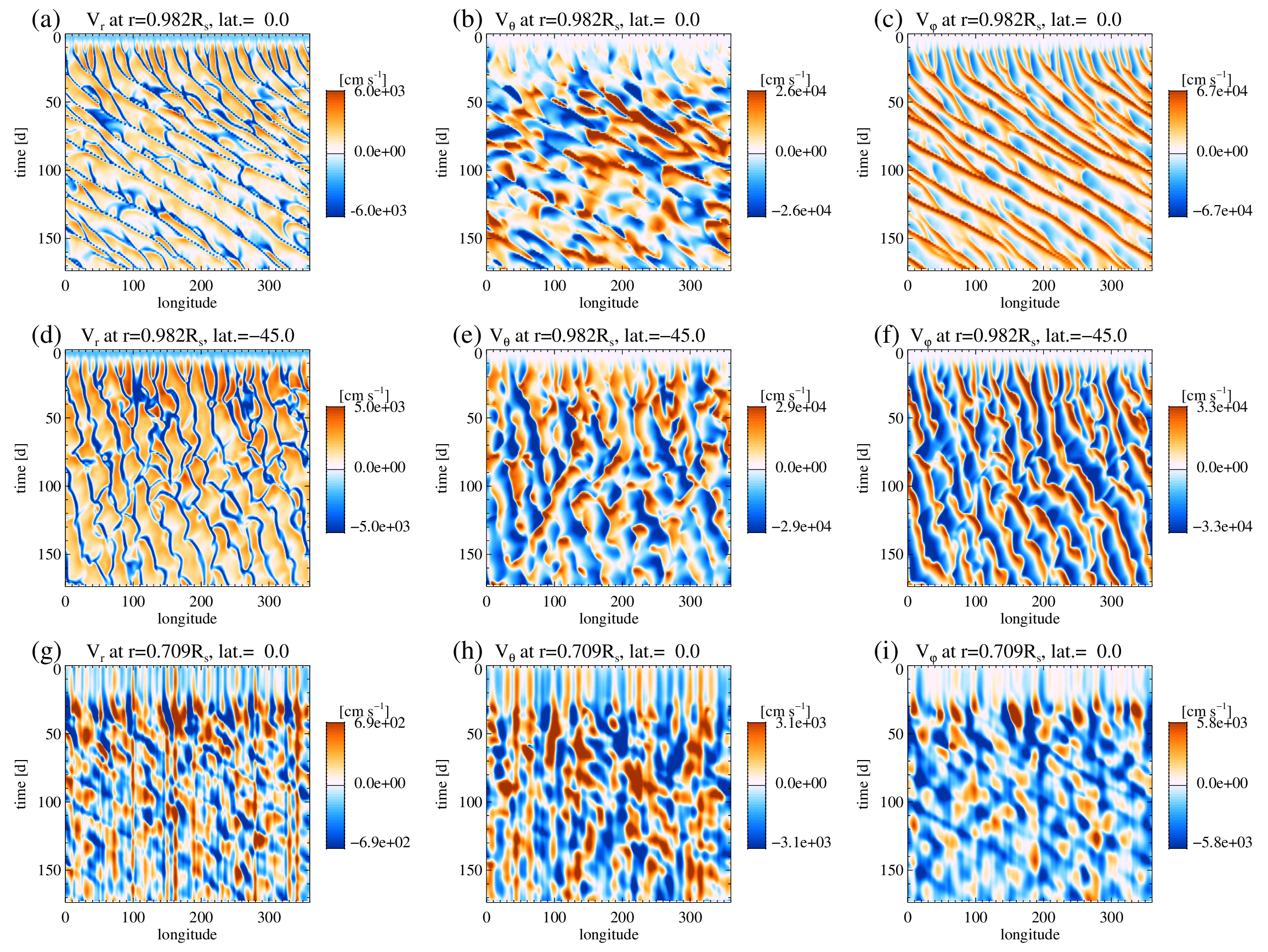}
\caption{%
The plasma velocity in the time-longitude plane (stack plot format).
In the top row (a) -- (c), the three components of the plasma velocity ($V_r$, $V_\theta$, $V_\phi$) sampled at $r=0.92R_\sun$ (near the top boundary surface) and on the equator (at the zero latitude) are shown.
In the middle (d) -- (f), the same quantities but sampled at 45 degrees south are shown.
In the bottom (g) -- (i), the velocity components sampled at $r=0.709R_\sun$ (near the bottom boundary surface) and on the equator are shown.
The time runs from top to bottom, and the longitude runs from left to right.
The prograde motions of the patterns are seen in the top row, and
the slightly weaker but noticeable prograde motions are seen in the middle row.
The colors are truncated at the 95th percentile of the absolute values of each variable.
}%
\label{fig11}
\end{figure}

\subsection{Long-term variations}
Figures \ref{fig10} and \ref{fig11} show the long-term trends of plasma flows in the convection zone. In Figure \ref{fig10}, the three components ($V_r$, $V_\theta$, and $V_\phi$) at three selected depths (0.982, 0.846, and 0.709 $R_\sun$) are shown in the time-latitude diagram
to capture the latitudinal motions of the plasma flow patterns.

An interesting point is found in the top row, where the patterns of the three components appear moving poleward at $r=0.982\,R_\sun$, even when the average direction of $V_\theta$ favors the equatorward motions (as seen in panel (e) of Figure \ref{fig06}).
This tendency becomes less noticeable in the middle of the convection zone,
and the equatorward motion of the patterns is found in the deeper part of the simulated convection zone (in the bottom row).

In Figure \ref{fig11}, the three components ($V_r$, $V_\theta$, and $V_\phi$)  are 
plotted on the longitude-time diagram.
The time here runs from top to bottom, following the convention of the stack plot format.
In the top row, the simulated plasma flows at the equator (noted as "lat=0" in the plot title).
The prograde motions of the patterns are clearly visible as the streaks running from top left to bottom right.
This is straightforwardly related to the solar-like differential rotation shown in the meridional section view in  Figure~\ref{fig06}(f).
In the middle row, the same properties, but sampled at 45 degrees South, are shown.
The prograde motions are still noticeable.
In the bottom row, the same quantities near the bottom boundary surface ($r= 0.709\,R_\sun$) are shown.
At this depth, no distinct tendency of the solar rotation can be found in part because of the boundary condition for the bottom boundary sphere.

\section{Discussion and Summary}\label{sctsummary}
We conducted a hydrodynamic simulation of the solar convection zone with the CHORUS++ code.
We used the values of solar luminosity and solar rotation rate, and the results are substantially different from the previous study \citep[][]{Chen23}.
One of the differences is the growth of convection, as indicated by the temporal profile of the kinetic energy shown in Figure \ref{fig03}.
In the previous study \citep[][]{Chen23}, the kinetic energy grows exponentially during the earliest phase, reaches a saturation state (or overshooting), then settles into a statistically stable equilibrium state.
In the presented model parameters, although we also observed exponential growth of the kinetic energy, the transition from the saturated state to the stable state was very gradual and slow.
Hence, this article analyzes the states that are still in transition.

Many of the parameters used in this study (tabulated in Table \ref{tblparam}) are not equal to those estimated for the Sun.
In particular, the parameters related to the fluid viscosity and the thermal conduction are set at orders of magnitude greater than the actual Sun.
We could use the kinematic viscosity 10 times smaller than in \citep[][]{Chen23}; however, the value in the solar convection zone is much smaller by several orders of magnitude.
Nonetheless, we here emphasize that the CHORUS++ code allows us to robustly conduct the simulations with relatively coarse grids.
The efficiency of the code will allow us to conduct the simulations
with the parameters much closer to the actual numbers of the solar convection zone, and better assess the dynamics of the solar interior, solar dynamo, and solar-cycle variations \citep[][and references therein]{Fan21}.

This article offers the model descriptions and preliminary analysis of the simulation results.
One item worth repeating here is the evaluation of the appropriateness of
the anelastic models.
As shown in Section \ref{sctdivrhovel}, the estimation least favoring the anelastic model still supports the divergence-free condition of mass flux and the sound-speed reduction method; in fact, the horizontal (latitudinal and longitudinal) gradients of the simulated mass density and temperature with the compressible HD code are very small with a deviation ratio of about $10^{-5}-10^{-4}$ relative to the average except for the region close to the solar surface.
This result overall is consistent with a study comparing the fully compressible simulations and the anelastic models \citep[][]{Verhoeven15} that found that the differences are overall small except for the regions with steep temperature gradients, like those at the outermost part of the solar convection zone.

Therefore, we find that the divergence-free constant-mass density assumption is appropriate for studying the dynamics of the deep solar convection zone.
At the same time, small gradients of the plasma properties are the primary drivers of the plasma motions; hence, the small-amplitude fluctuation of the plasma properties can affect the long-term dynamics.
An implication of this study is that the contribution of the near-surface density fluctuations must be considered when the model includes the shallow subsurface region, as the time scale there is estimated at an order of a day.
A critically important expansion of the convection model is to set the upper boundary surface as close to the solar surface ($r=1\,R_\sun$) with much higher resolution (smaller grid size) in the radial direction to capture the plasma gradients near the solar surface and properly simulate the most turbulent part of the Sun.

In the course of the transition of the convection system toward the equilibrium state, the simulated state exhibits a solar-like differential rotation pattern.
The extension of the simulation time is necessary to confirm that the simulated system will eventually reach solar or anti-solar rotation patterns \citep[e.g.,][]{Gilman77,Gastine14} and/or assess how long the presented patterns will last.
In the presented model setting, the magnetic field is not included, and the near-surface heat-escaping mechanism is modeled as a radiation dissipation term that depends on radius.
The results are expected to be more realistic when we include the magnetic field and MHD interactions in the convection zone \citep[e.g.,][]{Hotta22}.
The uniform heat flux is assumed at the bottom of the convection zone in the presented simulations.
This assumption can be modified to examine the influence of the horizontal (latitudinal and longitudinal) gradients in plasma properties on the long-term solar dynamo \citep[e.g.,][]{Bekki24}.
In the presented study, we set the radial component of the plasma flow to zero, $V_r=0$, on the boundary spheres together with the stress-free conditions for the horizontal components.
In the actual Sun, the penetration of convection flows into the radiation zone is possible and can strongly influence the entire dynamics of solar/stellar cycles \citep[e.g.,][]{Korre24}.
Furthermore, the physical processes acting in the radiative zone can substantially affect the tachocline and, hence, the convection zone.
For example, \citet{Matilsky22} conducted simulation studies for the region from 0.49 to 0.95\,$R_\sun$ to assess the influences of the radiation zone on the solar cycle, and \citet{Brun22} surveyed the influences in the sun-like stars with various values of the rotation rates and masses.
In the future, the interactions between the convective zone and the radiation zone across the tachocline will be assessed by setting the radius of the bottom boundary surface below 0.7\,$R_\sun$ with the CHORUS++ model.

Recently, a new MHD version of the CHORUS++ model has been intensely tested \citep[][]{Paoli25}.
We plan to modify and expand the model's capabilities to capture the dynamics of the solar interior realistically.
This new version will enable investigations of convection dynamics that are not yet fully understood.

\acknowledgments
The authors appreciate the anonymous referee for constructive comments and suggestions, which helped improve this article.
This work was supported by the NASA DRIVE Center COFFIES (Consequences Of Fields and Flows in the Interior and Exterior of the Sun) grant 80NSSC22M0162.
Computational resources supporting this work were provided by the NASA High-End Computing (HEC) Program through the NASA Advanced Supercomputing (NAS) Division at Ames Research Center.
C.L. thanks a National Science Foundation (NSF) award (No. 2310372).
%
%
\bibliography{reference}
\end{document}